\documentclass[pra,aps,twocolumn,showpacs]{revtex4}

\usepackage{bm}
\usepackage{amssymb}

\usepackage{graphicx,epsfig}

\begin{document}

\title{Linear optical Fredkin gate based on partial-SWAP gate}

\author{Jarom\'{\i}r Fiur\'{a}\v{s}ek} 
\affiliation{Department of Optics, Palack\'{y} University, 17. listopadu 50,
77200 Olomouc, Czech Republic}

\begin{abstract}
We propose a scheme for linear optical quantum Fredkin gate based on the combination of recently experimentally
demonstrated  linear optical partial SWAP gate and controlled-Z gates. Both heralded gate and simplified postselected gate operating in the coincidence basis are designed.
The suggested setups have a simple structure and require stabilization of only a single Mach-Zehnder interferometer. A proof-of-principle experimental demonstration of the postselected Fredkin gate appears to be feasible and within the reach of current technology.

\end{abstract}


\maketitle

\section{Introduction}

Linear optics quantum information processing has undergone a rapid development during last decade \cite{Kok07,Pan08}. 
While the suitability of photons for quantum communication has long been
recognized, it was initially thought that all-optical quantum computing is prohibited by the
lack of sufficiently strong nonlinear coupling between single photons.
However, Knill, Laflamme and Milburn showed in a seminal paper \cite{Knill01} that 
this obstacle is not fundamental. Two-qubit quantum
gates such as controlled NOT (CNOT)  can be implemented between two photons with certain probability using auxiliary single photons, passive linear optical interferometers, single
photon detectors and feedforward. If combined with quantum error correction, a universal
quantum computer could be constructed in this way, although the required overhead in resources
may be high.

This breakthrough was followed by many further theoretical developments which resulted in simplified schemes for the elementary two-qubit CNOT and CZ gates \cite{Ralph01,Pittman01,Hofmann02}, that have been subsequently demonstrated experimentally \cite{Pittman03,Brien03,Gasparoni04,Zhao05,Langford05,Kiesel05,Okamoto05,Bao07,Clark08,Politi08}. Moreover, special two-qubit fusion gates \cite{Browne05} were proposed for efficient generation of multiphoton cluster states that can be used as a resource for one-way quantum computing \cite{Raussendorf01,Nielsen04,Walther05,Prevedel07,Lu07}. 
In principle, arbitrary multiqubit quantum gate can be decomposed into a sequence of
CNOT gates and single-qubit gates. 
However, this approach to designing  more complex linear-optical quantum gates currently  faces a major difficulty, namely a scalability problem. Each heralded CNOT gate requires at
least two auxiliary photons. Since current experiments are limited to six-photon coincidence
detections \cite{Zhang06}, already concatenation of two CNOT gates appears to be a major experimental task. 

Realization of a generic two-qubit gate may require a sequence of up to three CNOTs
combined with single-qubit rotations. 
In case of three-qubit gates, the situation is even worse. For instance, the 
Fredkin or Toffoli gate require five two-qubit gates \cite{Smolin96}.
It is thus desirable to seek alternative simpler schemes for complex multi-qubit gates 
that are not based on sequence of CNOT gates.
Such approach recently proved to be very fruitful and lead to realization of several novel two-qubit linear optical gates.
Dedicated scheme for two-qubit partial SWAP gate operating in the coincidence basis
has been suggested and demonstrated experimentally \cite{Fiurasek08,Cernoch08}. Also  realization of arbitrary two-qubit controlled unitary gate has been reported very recently \cite{Lanyon08}. 

In the domain of three-qubit gates, attention was mainly paid to the fundamental Toffoli and Fredkin gates.
A scheme for linear optical Toffoli gate operating in the coincidence basis and
based on multiphoton interference has been put forward in Ref. \cite{Fiurasek06}. A different simpler scheme
for Toffoli gate has been described in Ref. \cite{Ralph07} and successfully experimentally demonstrated
soon afterwords \cite{Lanyon08}. This major achievement provides a strong motivation for the attempts to
realize also other three-qubit gates such as Fredkin gate. A scheme for heralded Fredkin gate
requiring altogether six ancilla photons has been suggested in Ref. \cite{Fiurasek06}. This scheme is inspired by the optical Fredkin gate proposed by Milburn \cite{Milburn89} but the cross-Kerr nonlinearity  is emulated using ancilla photons, interference and single-photon detection. The drawback of this first proposal is that it requires numerical design of a rather complicated interferometer. 
An alternative scheme for Fredkin gate has been very recently proposed by Gong, Guo and Ralph
(GGR) \cite{Gong08}. Their heralded Fredkin gate has a nice structure, but requires eight ancilla photons. They also propose a postselected Fredkin gate, which operates in the coincidence basis and
requires only a single pair of ancilla photons in maximally entangled state. However,
that scheme is still rather complex, with many optical elements and several
interferometers that would have to be stabilized simultaneously.

In the present paper we describe yet another scheme for both heralded and postselected Fredkin
gate. Recall that Fredkin gate acts as a controlled SWAP gate, the states of two target qubits are exchanged if and only if the control qubit is in computational basis state $|1\rangle$. We use the standard encoding of qubits into polarization states of single photons and horizontal and vertical polarization states $|H\rangle$ and $|V\rangle$ represent the computational basis states $|0\rangle$ and $|1\rangle$, respectively. Our design is based on the combination of the recently demonstrated linear optical partial SWAP gates \cite{Cernoch08} and CZ gates \cite{Kok07}. 
The resulting linear optical circuits have very simple structure. 
They involve less optical elements than previous proposals \cite{Fiurasek06,Gong08} 
and require stabilization of only a single Mach-Zehnder interferometer. As we shall argue below, proof-of-principle experimental demonstration of the postselected Fredkin gate is fully within the reach of present technology.

The rest of the paper is organized as follows. In Section II we describe the scheme for 
heralded linear optical quantum Fredkin gate. A simplified scheme for postselected Fredkin gate operating in the coincidence basis is presented in Section III. Finally, conclusions are drawn in Section IV.

\begin{figure}[!t!]
\centerline{\includegraphics[width=0.99\linewidth]{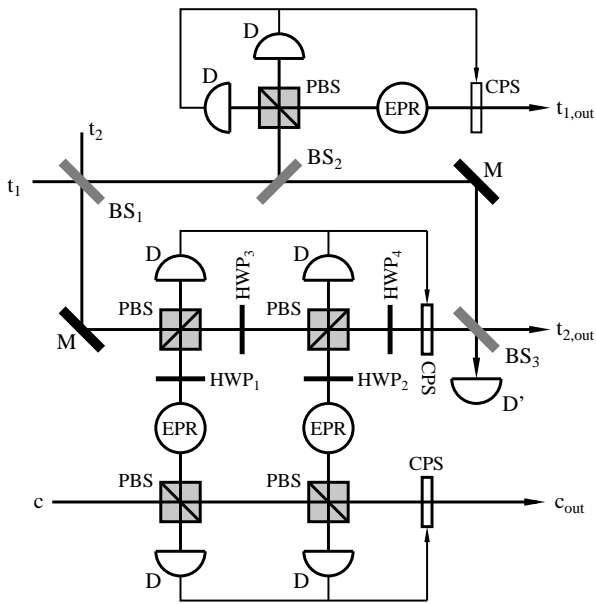}}
\caption{Linear optical Fredkin gate. BS$_1$ -- balanced beam splitter; 
BS$_2$ and BS$_3$ -- unbalanced beam splitters; 
M -- mirror; PBS -- polarizing beam
splitter that totally transmits horizontally polarized photons and reflects vertically
polarized photons; HWP -- half-wave plate; EPR -- source of auxiliary pair of entangled photons; CPS -- phase shifter controlled by electronic signal from the single photon detectors; D --
single-photon detection block performing polarization analysis in the diagonal linear
polarization basis; D' -- single-photon detector. The gate is successful if
each block D detects a single photon and detector D' detects no photons.}
\label{Fredkin-heralded}
\end{figure}

\section{Heralded Fredkin gate}

The heralded linear optical Fredkin gate based on the partial SWAP gate is shown in
Fig.~\ref{Fredkin-heralded}. The two target photons $t_1$ and $t_2$ propagate through a Mach-Zehnder (MZ) interferometer that consists of two beam splitters BS$_1$ and BS$_3$ and two mirrors M.
A fraction of the signal in the upper arm of the interferometer is reflected by the beam
splitter BS$_2$ and a quantum teleportation is used to verify in a quantum non-demolition
manner \cite{Kok02,Fiurasek06} that exactly a single photon is present in the output port $t_{1,\mathrm{out}}$. The other target photon should reach the output port $t_{2,\mathrm{out}}$ of the MZ interferometer. A single-photon detector D' is placed on the other output port to verify
that no photon leaked to that port. The gate succeeds only if $D'$ does not detect any
photon.

If the target photons are in symmetric two-qubit polarization state, then they bunch on the balanced beam splitter BS$_1$  
and both propagate either through the upper arm or through the lower arm of the interferometer. The photons reach the output ports $t_{1,\mathrm{out}}$ and $t_{2,\mathrm{out}}$ only if they both propagate through the upper arm,
one photon is reflected by BS$_2$, the other is transmitted through BS$_2$ and then reflected by BS$_3$. If the photons are initially in the antisymmetric singlet Bell state $|\Psi^{-}\rangle=\frac{1}{\sqrt{2}}(|H\rangle|V\rangle-|V\rangle|H\rangle)$, then the bunching does not occur,   the photons remain in the state $|\Psi^{-}\rangle$ after the interference on BS$_1$ and each photon propagates through one arm of the interferometer.

The fact that, effectively, a single target photon propagates through the lower interferometer arm only if the target photons are in the antisymmetric state is used to implement the controlled SWAP gate.
Note that the unitary two-qubit SWAP gate is diagonal in Bell state basis and the antisymmetric singlet Bell state acquires a $\pi$ phase shift with respect to the symmetric Bell states,
\[
U_{\mathrm{SWAP}}=\Pi_{+}-\Pi_{-},
\]
where $\Pi_{+}=\openone-\Pi_{-}$ and $\Pi_{-}=|\Psi^{-}\rangle\langle\Psi^{-}|$ are projectors onto the symmetric and antisymmetric subspaces of the Hilbert space of two qubits, respectively.
In our scheme, two heralded CZ gates are applied between the control and target photons. If the control photon is in state $|H\rangle$ then nothing happens but if it is in state $|V\rangle$ then the singlet Bell state of target photons is multiplied by $-1$ as required, 
\begin{equation}
\begin{array}{l}
|\Psi^{-}\rangle_{t_1t_2}|H\rangle_c\rightarrow  |\Psi^{-}\rangle_{t_1t_2}|H\rangle_c,
\\[2mm]
|\Psi^{-}\rangle_{t_1t_2}|V\rangle_c\rightarrow  -|\Psi^{-}\rangle_{t_1t_2}|V\rangle_c.
\end{array}
\label{psiminus}
\end{equation}
It is not difficult to verify that if the target photons are in an arbitrary symmetric state $|\Phi\rangle_{t_1t_2}$, then the optical circuit applies the identity transformation without any phase shift, $|\Phi\rangle_{t_1t_2}|\psi\rangle_c \rightarrow |\Phi\rangle_{t_1t_2}|\psi\rangle_c$.

\begin{figure}[!t!]
\centerline{\includegraphics[width=0.65\linewidth]{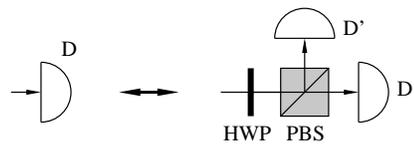}}
\caption{Structure of detection block for polarization analysis. Each block D consists of a half-wave plate (HWP) rotated by 22.5 degrees, polarizing beam splitter (PBS) and two single photon detectors D'.}
\label{Fredkin-Dblock}
\end{figure}

Let us now describe key parts of the circuit in more detail.
The QND measurement of single photon in the output port $t_{1,\mathrm{out}}$ is performed by quantum teleportation. One photon from an auxiliary pair of photons prepared 
in maximally entangled EPR state  $|\Phi^{+}\rangle=\frac{1}{\sqrt{2}}(|H\rangle|H\rangle+|V\rangle|V\rangle)$ 
interferes with the reflected signal on a polarizing beam splitter PBS. 
The output ports of the PBS are monitored by detection blocks D, each consisting of a
half-wave plate rotated by 22.5 degrees, polarizing beam splitter and two single-photon
detectors, see Fig.~\ref{Fredkin-Dblock}. The detection block D thus performs polarization measurement in the diagonal
basis $|\pm\rangle = \frac{1}{\sqrt{2}}(|H\rangle \pm |V\rangle)$. We assume that the detectors have unit
efficiency and can resolve number of photons in  the optical beam. If each detection block 
detects exactly a single photon then we know that a single photon was present in the beam
reflected by BS$_2$ and the polarization state of this photon has been teleported onto the remaining photon from the EPR pair. If the measurement results have different parity (i.e. if the outcomes read $+-$ or $-+$) then we need to apply a corrective
unitary transformation $\sigma_Z= |H\rangle\langle H| -|V\rangle\langle V |$ to the output photon.
This can be accomplished by a fast electrooptical modulator controlled by the signal from
the detectors \cite{Sciarrino06,Prevedel07}, as indicated in the figure. The feedforward can be avoided if one selects only the coincidence events
where both detection blocks D yield identical measurement outcomes $++$ or $--$.
This would simplify the scheme but reduce the success probability of the gate by factor
$\frac{1}{2}$.

The scheme further involves two  controlled-Z  gates between  the control photon c
and a target photon $t_2$ propagating through the lower arm of the interferometer. 
The CZ gate applies the following unitary transformation,
\begin{equation}
\begin{array}{lcl}
|HH\rangle \rightarrow |HH\rangle, & \qquad &
|HV\rangle \rightarrow |HV\rangle, \\
|VH\rangle \rightarrow |VH\rangle, & \qquad &
|VV\rangle \rightarrow -|VV\rangle.
\end{array}
\end{equation}
 Each linear-optical heralded CZ gate \cite{Pittman01,Pittman03,Gasparoni04,Zhao05} consists of 
an auxiliary EPR pair of photons in Bell state $|\Phi^{+}\rangle$, two polarizing beam
splitters PBS, a half-wave plate (HWP$_1$ and HWP$_2$, respectively) rotated by 22.5 degrees,
and two detection blocks D. The CZ gate succeeds if each block D detects exactly a single
photon.  A unitary operation depending on the measurement outcomes has to be applied to the output photons, as schematically indicated in Fig.~\ref{Fredkin-heralded}. 
 The HWP$_3$ and HWP$_4$ are rotated by 45 degrees such that
they exchange the vertical and horizontal polarization, 
\begin{equation}
|H\rangle\rightarrow |V\rangle, \qquad |V\rangle\rightarrow |H\rangle.
\end{equation}
The net result of combination of the CZ gates and half-wave plates is the application
of a $\pi$ phase shift to the target photon in the lower interferometer arm if the control photon is vertically polarized.

The success probability of the gate for input antisymmetric state of target photons
$|\Psi^{-}\rangle_{t_1 t_2}$ can be calculated as,
\begin{equation}
P_{-}=R_2 \times \frac{1}{2} \times \frac{1}{4} \times \frac{1}{4} \times T_3,
\end{equation}
where $T_j$ is the intensity transmittance of jth beam splitter and $R_j=1-T_j$ is the intensity reflectance, $\frac{1}{2}$ is the success probability of teleportation-based QND measurement, and 
$\frac{1}{4}$ is the success probability of each CZ gate. 
The success probability for a symmetric input state can be calculated in a similar manner
and we obtain,
\begin{equation}
P_{+}=\frac{1}{2}\times 2T_2 R_2 \times \frac{1}{2} \times \frac{1}{4}\times \frac{1}{4} \times R_3 .
\end{equation}
Here the first factor $\frac{1}{2}$ is the probability that after interference on BS$_1$ both target photons will propagate through the upper interferometer arm, $2T_2R_2$  is the probability that one photon is reflected by BS$_2$ and one is transmitted through BS$_2$, and $\frac{1}{4}$ is the  success probability of the CZ gate if there is no target photon in the lower arm of the interferometer and the gate acts as identity on the control qubit.

A unitary Fredkin gate will be conditionally implemented provided that the success
probability for the symmetric and anti-symmetric states of target photons will be identical,
\begin{equation}
P_{+}=P_{-}.
\end{equation}
This condition implies dependence between the transmittances of BS$_2$ and BS$_3$,
\begin{equation}
T_2=\frac{T_{3}}{1-T_3}.
\end{equation}
We can optimize the transmittance $T_3$ such as to maximize the overall success probability
of the Fredkin gate,
\begin{equation}
P_{}=\frac{1}{32}\frac{1-2T_3}{1-T_3}T_3.
\label{Ptot}
\end{equation}
The optimization yields $T_{3,\mathrm{opt}}=1-1/\sqrt{2}$ and $T_{2,\mathrm{opt}}=\sqrt{2}-1$.
On inserting back into Eq. (\ref{Ptot}) we obtain,
\begin{equation}
P_{\mathrm{max}}=\frac{3-2\sqrt{2}}{32} \approx 5.36 \times 10^{-3}.
\end{equation}

Let us summarize the most important features of the Fredkin gate. The gate requires 
six auxiliary photons forming three maximally entangled EPR pairs. The success of the gate
is heralded by a detection of single photon by each of the six detection blocks D and 
simultaneously the detector D' should not detect any photon. On success, the Fredkin gate
is applied to polarization states of the photons and the photons are available at the
output for further processing. The success of the gate can be maximized by optimizing
transmittances of the beamsplitters in the setup. The maximum success probability
$P_{\mathrm{max}}\approx 5.36 \times 10^{-3}$ is about $5.5$ times higher than the success
probability of the Fredkin gate proposed by GGR \cite{Gong08}. Moreover, the present scheme requires
only six ancilla photons while the GGR scheme involves eight ancilla photons (which are
needed to implement four heralded CNOT gates).  In comparison to the linear optical
Fredkin gate proposed  in  Ref. \cite{Fiurasek06}, the present scheme has much simpler and cleaner 
structure and its design is fully analytical.

\section{Fredkin gate operating in coincidence basis}

In current experiments, time-synchronized correlated or entangled photon pairs are generated by means of spontaneous  parametric down-conversion (SPDC) in nonlinear crystals pumped by femtosecond lasers. 
The experimental demonstration of the heralded Fredkin gate would require simultaneous
generation of nine photons and detection of nine-photon coincidences. This is a formidable
 task mainly because the probability of generation and detection of $N$ photons
in experiments using SPDC sources decreases exponentially with growing $N$.

\begin{figure}[!t!]
\centerline{\includegraphics[width=0.99\linewidth]{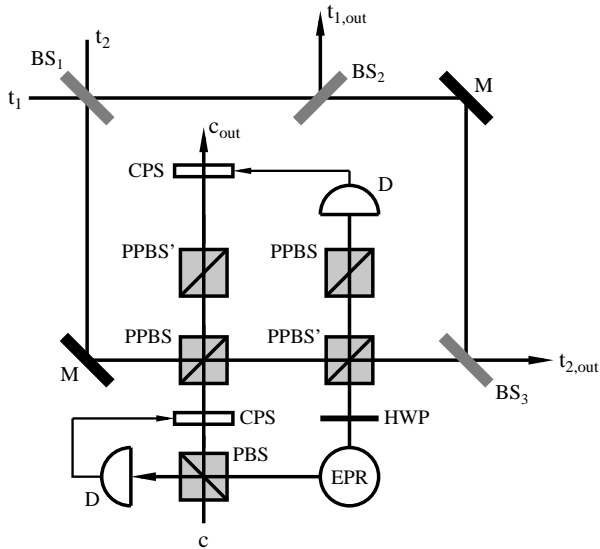}}
\caption{Linear optical Fredkin gate operating in the coincidence basis. PPBS -- partially
polarizing beam splitter with transmittances $T_H=1$ and $T_V=\frac{1}{3}$; PPBS' --
partially polarizing beam splitter with transmittances $T_H^\prime=\frac{1}{3}$ and $T_V^\prime=1$.
Other labels have the same meaning as in Fig.~\ref{Fredkin-heralded}. The gate succeeds if each detection block D
detects a single photon and if a single photon is detected in each output port of the gate.}
\label{Fredkin-coincidence}
\end{figure}

The complexity of the Fredkin gate can be significantly reduced if we design a gate that 
operates in the coincidence basis \cite{Ralph02}. This means that the success of the gate is heralded by a 
detection of a single photon in each output port of the gate. The photons are thus consumed
and not available for further processing unless their detection is carried out in a quantum
non-demolition manner. The resulting scheme for Fredkin gate operating in the coincidence
basis  is depicted in Fig.~\ref{Fredkin-coincidence}. The basic structure of the scheme is the same as that of the heralded Fredkin gate in Fig.~\ref{Fredkin-heralded} but there are several important changes and simplifications.

First, the teleportation block in the output port  t$_{1,\mathrm{out}}$ is removed because it is not
necessary for a gate operating in the coincidence basis. 
Second, the two heralded CZ gates between the control photon and the target photon propagating
through the lower arm of the Mach-Zehnder interferometer were replaced by two CZ gates
operating in the coincidence basis \cite{Langford05,Kiesel05,Okamoto05}. A single auxiliary EPR pair and a quantum parity check \cite{Pittman01}
are used to encode the input state of the control photon onto an entangled state of two
photons. The combination of PBS, detection block D, controlled phase shifter CPS and half-wave
plate HWP rotated by 45 degrees, transforms with probability $\frac{1}{2}$ 
the input state of control photon $\alpha|H\rangle_c+\beta|V\rangle_c$ onto a state
$\alpha|H\rangle_c|V\rangle_{\mathrm{aux}}+\beta|V\rangle_c|H\rangle_{\mathrm{aux}}$.

The CZ gate between the control and target photon is implemented by a sequence of two
partially polarizing beam splitters PPBS and PPBS'. The beam splitter PPBS  totally transmits
horizontally polarized photons and transmits vertically polarized photons with probability
$T_V=\frac{1}{3}$.  The beam splitter PPBS' perfectly transmits vertically polarized photons and
transmits horizontally polarized photons with probability $T_H'=\frac{1}{3}$. The CZ gate between
the auxiliary photon and the target photon is similarly
realized using PPBS' and PPBS. The two CZ gates succeed if a single photon is detected by the
upper detection block D, a single control photon is present in the output port c$_{\mathrm{out}}$, and the 
target photon reaches the output port t$_{2,\mathrm{out}}$. 

The encoding of the state of control photon into two-photon state is removed by quantum
erasing, which is performed by the detection block D that measures the auxiliary photon in
diagonal polarization basis. If the measurement outcome is $|-\rangle$, then a correcting
unitary $\sigma_Z$ operation has to be applied to the output control photon. 

Let us now evaluate the probability of success of the scheme for symmetric and antisymmetric
input state of the target photons, respectively. Consider first symmetric input state.
The probability that the target photons reach the output ports reads $T_2 R_2 R_3$.
Note that a sequence of PPBS and PPBS' acts
as a grey filter with transmittance $\frac{1}{3}$ that equally attenuates all polarization states.
Therefore if no target photon travels through the lower interferometer arm, the control photon
is transferred to the output without any change of its polarization state. The event that
each detection block D detects a single photon and that the control photon is present in the
output port c$_{\mathrm{out}}$ occurs with probability $\frac{1}{2}\times \frac{1}{3}\times
\frac{1}{3}=\frac{1}{18}$. The total success probability for symmetric input in target ports
is thus given by,
\begin{equation}
P_{+,\mathrm{coinc}}=\frac{1}{18}T_2R_2 R_3.
\end{equation}
Consider now the antisymmetric input state $|\Psi^{-}\rangle_{t_1t_2}$. One target photon
travels through each arm of the interferometer. One target photon is reflected by BS$_2$ and the other
photon is subject to two CZ gates and then is transmitted through BS$_3$. The gate 
succeeds with probability
\begin{equation}
P_{-,\mathrm{coinc}}=\frac{1}{54} R_2 T_3.
\end{equation}
The net effect of the two CZ gates is that the target photon acquires a phase shift $\pi$ if the
control photon is in state $|V\rangle$, i.e. the transformation (\ref{psiminus}) is implemented.
The scheme realizes unitary Fredkin gate if the success probabilities are the same for symmetric and
antisymmetric outputs, $P_{+,\mathrm{coinc}}=P_{-,\mathrm{coinc}}$. This yields,
\begin{equation}
T_{2}=\frac{T_3}{3(1-T_3)}.
\end{equation}
Similarly as before we can optimize $T_3$ so as to maximize the total probability of success.
We obtain $T_{3,\mathrm{opt}}=\frac{1}{2}$ and $T_{2,\mathrm{opt}}=\frac{1}{3}$, which yields
\begin{equation}
P_{\mathrm{\mathrm{max,coinc}}}=\frac{1}{162}.
\end{equation}
This is slightly higher than the success probability $1/192$ 
of the Fredkin gate operating in the coincidence basis proposed by GGR \cite{Gong08}.

The Fredkin gate in Fig.~\ref{Fredkin-coincidence} succeeds if a single photon is detected in each output port of the
gate and if simultaneously each detection block D detects a single photon. 
Experimental demonstration of this gate would therefore require generation and detection of
five photons, which is within reach of present technology. Moreover, for a
proof-of-principle demonstration the scheme can be further simplified. In particular, one could
avoid the encoding of the input state of control photon onto a photon pair by quantum parity check and instead
directly use a photon pair produced by SPDC in entangled polarization state.
The experiment would then boil down to simultaneous generation of two photon pairs
by SPDC and four-photon coincidence detection, which is well mastered experimentally. 
Let $p_t$ denote the probability of generation of a pair of  target photons and let $p_c$ denote the probability of generation of the entangled photon pair representing encoded control qubit.
With probability $p_t p_c$ we get the required four-photon input. However, with probability $\approx p_t^2$
two pairs of target photons are produced which can result is false four-photon coincidence events.
On the other hand, if two photon pairs encoding control qubit are produced, then no photon will be registered in the output port t$_{1,\mathrm{out}}$, so these events would not contribute to observed four-fold coincidences.
We therefore have to compare the probabilities of correct events $p_t p_c$ and wrong events $p_t^2$.
The unwanted events when two pairs of photons are injected into the target input ports and no photons are injected into control input ports  can be suppressed if $p_t \ll p_c \ll 1$. This can be achieved, e.g., by adjusting the intensity of the pump laser beams. 

\section{Conclusions}

In summary, we have proposed a scheme for linear optical Fredkin gate that is based on the
combination of the linear optical partial SWAP gate and linear optical CZ gates. 
It is worth emphasizing that both these elementary two-qubit gates have been recently
successfully demonstrated experimentally. An important and distinct feature of the present
scheme is that it requires stabilization of only a single Mach-Zehnder 
interferometer, which is well mastered experimentally \cite{Cernoch08}. A proof-of-principle
demonstration of the Fredkin gate operating in the coincidence basis requires simultaneous
generation of two correlated photon pairs and detection of
four-photon coincidences, which is fully achievable with current technology. 
The present proposal thus represents an important step towards experimental realization 
of the linear optical Fredkin gate.

\begin{acknowledgments}
This work was supported by Research Projects  ``Center of Modern
Optics'' (LC06007) and ``Measurement and Information in Optics''
(MSM 6198959213) of the Czech Ministry of Education and by GACR (Grant No.
202/08/0224).

\end{acknowledgments}


\begin{thebibliography}{99}

\bibitem{Kok07}
P. Kok, W.J. Munro, K. Nemoto, T.C. Ralph, J. P. Dowling, and 
G.J. Milburn, Rev. Mod. Phys. \textbf{79}, 135 (2007).



\bibitem{Pan08}
J.-W. Pan, Z.-B. Chen, M. Zukowski, H. Weinfurter, and A. Zeilinger,
arXiv:0805.2853.





\bibitem{Knill01}
E. Knill, R. Laflamme, and G.J. Milburn,
Nature (London) \textbf{409}, 46 (2001). 


\bibitem{Ralph01}
T.C. Ralph, A.G. White, W.J. Munro, and G.J. Milburn, 
Phys. Rev. A \textbf{65}, 012314 (2001).



\bibitem{Pittman01}
T.B. Pittman, B.C. Jacobs, J.D. Franson,
 Phys. Rev. A \textbf{64}, 062311 (2001). 


\bibitem{Hofmann02}
H.F. Hofmann and S. Takeuchi,
Phys. Rev. A \textbf{66}, 024308 (2002).


\bibitem{Pittman03}
T. B. Pittman, M. J. Fitch, B. C Jacobs, and J. D. Franson,
 Phys. Rev. A \textbf{68}, 032316 (2003).


\bibitem{Brien03}
J. L. O'Brien, G. J. Pryde, A. G. White, T. C. Ralph, and D. Branning,
Nature (London) \textbf{426}, 264 (2003).



\bibitem{Gasparoni04}
S. Gasparoni, J.-W. Pan, P. Walther, T. Rudolph, and A. Zeilinger,
Phys. Rev. Lett. \textbf{93}, 020504 (2004).  


\bibitem{Zhao05}
Z.~Zhao, A.~N. Zhang, Y.~A. Chen, H.~Zhang, J.~F. Du, T.~Yang, and J.~W. Pan,
 Phys. Rev. Lett., {\bf 94} 030501, 2005.



\bibitem{Langford05}
N. K. Langford, T. J. Weinhold, R. Prevedel, K. J. Resch, A. Gilchrist, J. L. O'Brien, 
G. J. Pryde, and A. G. White, 
Phys. Rev. Lett. \textbf{95}, 210504  (2005).


\bibitem{Kiesel05}
N. Kiesel, Ch. Schmid, U. Weber, R. Ursin, and H. Weinfurter,
Phys. Rev. Lett. \textbf{95}, 210505  (2005).


\bibitem{Okamoto05}
R. Okamoto, H. F. Hofmann, S. Takeuchi, and K. Sasaki,
Phys. Rev. Lett. \textbf{95}, 210506  (2005).


\bibitem{Bao07}
X.-H. Bao, T.-Y. Chen, Q. Zhang, J. Yang, H. Zhang, T. Yang, and J.-W. Pan,
Phys. Rev. Lett. \textbf{98}, 170502 (2007). 



\bibitem{Clark08}
A. S. Clark, J. Fulconis, J. G. Rarity, W. J. Wadsworth, and J. L. O'Brien,
arXiv:0802.1676.


\bibitem{Politi08}
A. Politi, M. J. Cryan, J. G. Rarity, S. Yu, and J. L. O'Brien,
Science \textbf{318}, 1567 (2008).


\bibitem{Browne05}
D.E. Browne and T. Rudolph, Phys. Rev. Lett. \textbf{95},
010501 (2005).


\bibitem{Raussendorf01}
R. Raussendorf and H.J. Briegel, Phys. Rev. Lett. \textbf{86},
5188 (2001).

\bibitem{Nielsen04}
M. A. Nielsen, Phys. Rev. Lett. \textbf{93}, 040503 (2004).


\bibitem{Walther05}
P. Walther, K. J. Resch, T. Rudolph, H. Weinfurter, V. Vedral, M. Aspelmeyer, 
and A. Zeilinger, Nature \textbf{434}, 169 (2005).

\bibitem{Lu07}
C.-Y. Lu, X.-Q. Zhou, O. G\"{u}hne, W.-B. Gao, J. Zhang, Z.-S. Yuan, A. Goebel, 
T. Yang, and J.-W. Pan, Nature Phys. \textbf{3}, 91 (2007). 


\bibitem{Prevedel07}
R. Prevedel, P. Walther, F. Tiefenbacher, P. Bohl, R. Kaltenbaek, T. Jennewein, 
and A. Zeilinger, Nature (London)  \textbf{445}, 65 (2007).  


\bibitem{Zhang06}  
Q. Zhang, A. Goebel, C. Wagenknecht, Y.-A. Chen, B. Zhao, T. Yang, A. Mair, 
J. Schmiedmayer and J.-W. Pan, Nature Phys. \textbf{2}, 678 (2006).


\bibitem{Smolin96}
J. A. Smolin and D. P. DiVincenzo, Phys. Rev. A \textbf{53}, 2855 (1996).


\bibitem{Fiurasek08}
J. Fiur\'{a}\v{s}ek and N.J. Cerf,  Phys. Rev. A \textbf{77}, 052308 (2008).



\bibitem{Cernoch08}
A. \v{C}ernoch, J. Soubusta, L. Bartu\v{s}kov\'{a}, M. Du\v{s}ek, and J. Fiur\'{a}\v{s}ek,  
Phys. Rev. Lett. \textbf{100}, 180501 (2008). 



\bibitem{Lanyon08}
B. P. Lanyon, M. Barbieri, M. P. Almeida, T. Jennewein, T. C. Ralph, 
K. J. Resch, G. J. Pryde, J. L. O'Brien, A. Gilchrist, and A. G. White, 
arXiv:0804.0272.



\bibitem{Fiurasek06} 
J. Fiur\'{a}\v{s}ek, Phys. Rev. A \textbf{73}, 062313 (2006).



\bibitem{Ralph07}
T.C. Ralph, K.J. Resch, and A. Gilchrist,
Phys. Rev. A \textbf{75}, 022313 (2007).



\bibitem{Milburn89}
G. J. Milburn, Phys. Rev. Lett. \textbf{62}, 2124 (1989). 



\bibitem{Gong08}
Y.-X. Gong, G.-C. Guo, and T. C. Ralph, 
 Phys. Rev. A \textbf{78}, 012305 (2008).



\bibitem{Kok02}
P. Kok, H. Lee, and J. P. Dowling,
Phys. Rev. A \textbf{66}, 063814 (2002); quant-ph/0202046v1.



\bibitem{Sciarrino06}
F. Sciarrino, M. Ricci, F. De Martini, R. Filip, and L. Mi\v{s}ta Jr.,
Phys. Rev. Lett. \textbf{96}, 020408 (2006). 








\bibitem{Ralph02}
T. C. Ralph, N. K. Langford, T. B. Bell, and A. G. White,
Phys. Rev. A \textbf{65}, 062324 (2002). 


\end{thebibliography}
\end{document}